# Driving Droplets by Curvi-Propulsion

Cunjing Lv[1,2], Chao Chen[1,2], Yin-Chuan Chuang[3], Fan-Gang Tseng[3,4], Yajun Yin[1],

Francois Grey[2], Quanshui Zheng[1,2*]

[1] Department of Engineering Mechanics, Tsinghua University, Beijing 100084, China

[2] Center for Nano and Micro Mechanics, Tsinghua University, Beijing 100084, China

[3] Department of Engineering and System Science, National Tsing Hua University, Hsinchu 30013, Taiwan

[4] Research Center for Applied Sciences, Academia Sinica, Taipei 11529, Taiwan

**How to make small liquid droplets move spontaneously and directionally on solid surfaces is a challenge in lab-on-chip technologies[1,2], DNA analysis[3], and heat exchangers[4]. The best-known mechanism, a wettability gradient, does not move droplets rapidly enough for most purposes and cannot move droplets smaller than a critical size defined by the contact angle hysteresis[5]. Here we report on a mechanism using curvature gradients, which we show is particularly effective at accelerating small droplets, and works for both hydrophilic and hydrophobic surfaces. Experiments for water droplets on glass cones in the sub-millimeter range show a maximum speed of 0.28 m/s, two orders of magnitude higher than obtained by wettability gradient. From simple considerations of droplet surface area change, we show that the force exerted on a droplet on a conical surface scales as the curvature gradient. This force therefore diverges for small droplets near the tip of a cone. We illustrate this using molecular dynamics simulations, and describe nanometer-scale droplets moving spontaneously at over 100 m/s on nano-cones.**

Most observations of the spontaneous motion of droplets on solid surfaces are variants of the Marangoni effect, due to wettability gradients[6-9]. Many techniques, such as thermal[4,7], chemical[2], electrochemical[10], and photochemical methods[11,12], can be used to make a flat surface possess continuously varying liquid contact angle. Droplets on such surfaces tend to move toward the region with lower contact angle. Although spontaneous motion has also been observed for droplets on substrates with uniform surface energy but varying surface roughness[13], its mechanism can still be categorized as a Marangoni effect because of the varying effective contact angle. The speed of spontaneous droplet motion resulting from such effects is typically in the range of micrometers to millimeters per second[4], far too low for applications in areas such as liquid-based thermal management of fuel cells and semiconductor devices.

The main obstacle to droplet motion on a solid surface arises from contact angle hysteresis. In order to overcome this difficulty, large droplets are typically used and external sources of energy are introduced such as vibrating or heating the surface[4,14,15]. Spontaneous motion up to 0.3 m/s was observed[4] for condensation droplets coming from saturated steam (100°C) on a silicon wafer with a radial gradient of surface energy. The current record for spontaneous droplet motion under ambient conditions is about 0.5 m/s[16,17], using a chemically patterned and nanotextured surface.

---

[*] To whom correspondence should be addressed. E-mail: zhengqs@tsinghua.edu.cn

Here we report a mechanism that can result in ultrafast spontaneous motion of micro- and nanoscale water droplets. In contrast with the Marangoni effect, neither surface energy gradient nor surface roughness variation is required. The purely geometrical consequences of the curvature gradient along any meridian of a cone provide the driving mechanism. Curvature-gradient-driven motion of liquids has been observed in specific circumstances, such as for wetting droplets on a conical surface18. What distinguishes the results described here from the observations and analyses reported earlier [2,5,18] is: (1) We show the mechanism, which we call curvi-propulsion, is valid not only for hydrophilic surfaces, but also for hydrophobic ones; (2) drops much smaller than considered previously can move spontaneously due to curvi-propulsion; (3) at the nanoscale, we predict unprecedented speeds for the motion, and confirm this by molecular dynamics simulations.

First, we will present the experimental and simulation-based evidence for the effect. Then we will explain these results using a simple model, and show that it accounts quantitatively for both the increase of the velocity at the nanoscale, and the persistence of the effect on hydrophobic surfaces.

To observe the effect experimentally, tapered surfaces as illustrated in the Insert of Fig. 1b were sharpened from glass tubes of 1.5 mm in diameter by a flaming/brown micropipette puller (Model P-1000, Sutter Instrument). The video frames shown in Fig. 1a illustrate a typical spontaneous motion of a 1 mm$^3$ water drop under ambient conditions on the tapered surface. The droplet moves toward the direction of increasing cross-section radius, $R$, of the tapered tube. The velocity, $v$, of the droplet is measured as a function of $R$ along the tapered tube, plotted as green triangles in Fig. 1b as the surface had been cleaned by deionized water. For this surface, the water contact angle, $\theta$, and the contact angle hysteresis, $\Delta\theta$, defined as the difference between the advancing and receding contact angles of the droplet, are measured as about 41° and 4°. Other data in Fig. 1b show the velocity of identical droplets on the same cone after treating its surface in an oxygen plasma (blue circles, with $\theta \approx 5°$ and $\Delta\theta \approx 2°$) and then in MTS solution (red squares, with $\theta \approx 0°$ and $\Delta\theta \approx 1.5°$). The velocity is non monotonic along the surface, but it can reach well over 0.2 m/s and it increases with decreasing contact angle. The maximum speed (~0.28 m/s) observed in these experiments is two orders of magnitude higher than those obtained by standard wettability gradients[5].

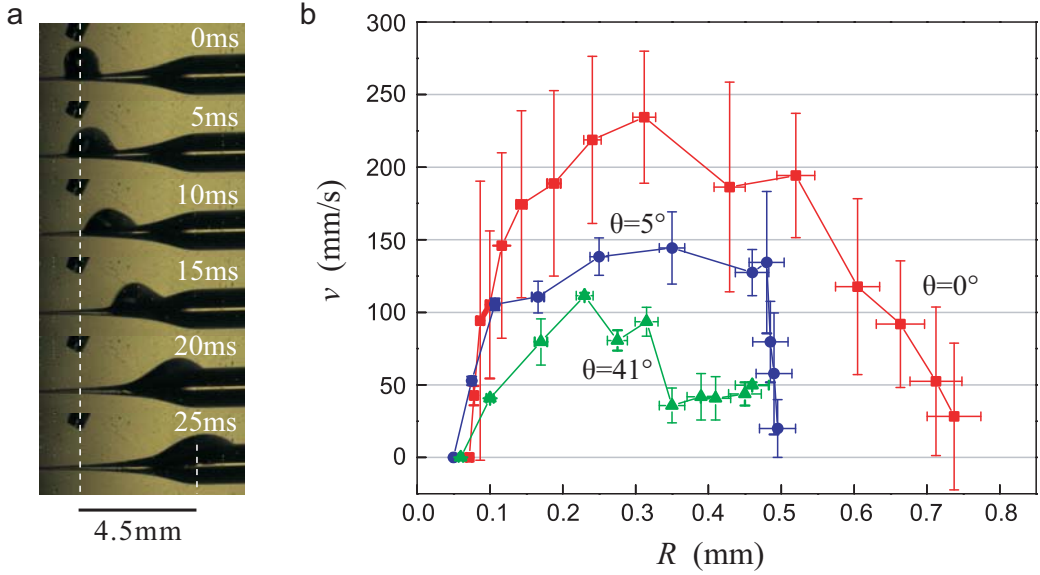

**Figure 1.** Experimentally observed fast spontaneous motion of 1 mm³ water droplets on tapered glass rods with different surface conditions. **a,** Video frames of a droplet moving on an untreated tapered glass surface with radius range 0.07 - 0.75 mm, contact angle $\theta \approx 41°$, and hysteresis $\Delta\theta \approx 4°$. **b,** Velocity, $v$, as a function of local radius, $R$, of the glass tapered tube, for droplets moving on the same tapered surface with three different surface conditions: untreated; O$_2$ plasma-treated ($\theta \approx 5°$, $\Delta\theta \approx 2°$); and MTS nanotextured plus O$_2$ plasma-treated ($\theta \approx 0°$, $\Delta\theta \approx 1.5°$) (see the videos in the Supplementary Information for more details).

We complemented these experiments with molecular dynamics (MD) simulations based on the LAMMPS platform[19], allowing us to describe similar phenomena at a nanoscale. A nano-cone was modeled as a rigid framework of a monolayer of conically rolled graphene, as shown in Figs. 2a and 2c. The droplet was modeled by a standard code (SPC/E)[20] with the same parameters as given in Ref. 21. A Lennard-Jones potential, $\phi(r) = 4\varepsilon[(\sigma/r)^{12} - (\sigma/r)^6]$, was used to characterize the cone-liquid van der Waals interaction[22], where $r$ denotes the distance between atoms. By fixing the equilibrium distance $\sigma$ at 0.319 nm and adopting different values 5.85 and 1.95 meV for the well depth $\varepsilon$, we obtained two different contact angles $\theta = 51°$ (hydrophilic) and 138° (hydrophobic). The model cone has a half-apex angle $\alpha = 19.5°$ and a height of 7 nm. As illustrated in Figs. 2a and 2c, we then cut off the tip at the heights of 1.5nm and 3.5nm, respectively, for modeling the motion of a 2nm-diameter droplet containing 339 molecules. We simulate motion on both the outer and inner conical surfaces. In the simulations, the temperature is kept at 300 K with a Nosé/Hoover thermostat, and the whole system is located in a finite vacuum box. Initially, the mass center of the droplet is fixed at a starting point for 100 ps, to reach thermal equilibrium. Then the droplet is released to move freely along the conical surface for the next 300 ps.

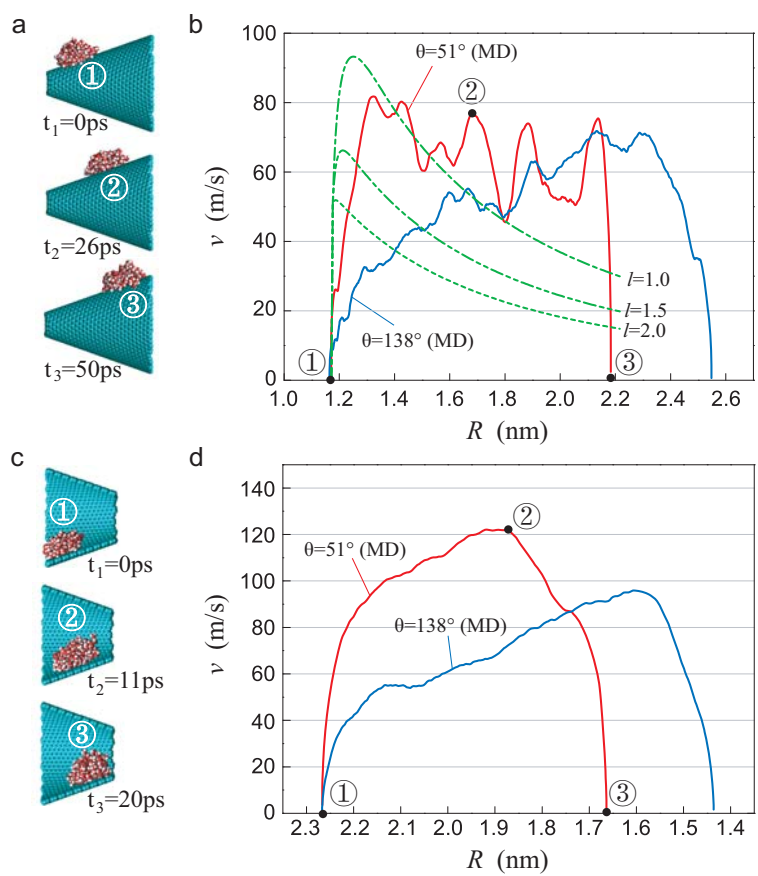

**Figure 2.** Molecular-dynamics simulations of spontaneous motion of a droplet consisting of 339 water molecules on the outer surface **a**, **b** and inner surface **c**, **d** of a nano-cone. In **b** the velocity of the droplet, $v$, which starts near the narrow end of the cone, is plotted as a function of the cross-section radius of the cone, $R$, until it reaches the large end of the cone, where it briefly stops. The red (solid) and blue (dot-dashed) correspond to water contact angles of 51° and 138° respectively. In **c** the corresponding velocity of the droplet moving inside the cone is plotted, from a starting point near the large end. Illustrations in **a**, **c** show droplet positions at three different times, $t_1$, $t_2$ and $t_3$, that are noted in the graphs (from the MD simulation movie, see Supplementary Information).

When the droplet is released near the small end of the external surface, it starts to move spontaneously toward the larger end, as illustrated in Figs. 2a and 2b, with the maximum velocity reaching briefly over 80 m/s, until the droplet encounters the large end of the cone, where it bounces off the potential barrier. The red (solid) curve in Fig. 2b depicts the dependence of velocity, $v$, on radius of the cone $R$ measured directly below the center of mass of the droplet, for contact angle $\theta = 51°$. Motion is shown up until the point where the droplet instantaneously comes to rest at the potential barrier. Snapshots of the droplet at three times during this motion are illustrated in Fig. 2a. Similar spontaneous motion is observed for the hydrophobic case ($\theta = 138°$) a maximum velocity of 70 m/s being reached more slowly than in the hydrophilic case. When the droplet is released on the inner surface, we observe it to move spontaneously towards the smaller end of the cone, as illustrated in Figs. 2c and 2d. Even higher maximum velocities are reached in this case, of over 120 m/s for a hydrophilic interaction and over 90 m/s for the hydrophobic case.

To the best of our knowledge, this is the first report of a mechanism driving spontaneous droplet motion on hydrophobic surfaces, an effect normally observed only on hydrophilic surfaces[18,23,24]. Most remarkable, however, are the speeds attained by the nanodroplets, which are several orders of magnitude higher than any prior observations or simulations. We note that nanodroplets will also move randomly due to thermal energy. But this effect is relatively small, corresponding at room temperature to undirected motion at about 20 m/s for the size of droplet we are simulating[25].

To explain this ultra-fast transport, we first explore how the surface energy of a droplet varies on a curved surface. For droplets that have diameters smaller than the capillary length, $\lambda_c = (\gamma/\rho g)^{1/2}$, where $\gamma$ and $\rho$ are the free surface tension and mass density of the liquid, and $g$ is the gravity acceleration, we can ignore the gravity potential. Thus, the total free energy can be quantified as[9,26]

$$U = \left( A_{LV} - A_{LS} \cos\theta \right) \tag{1}$$

where $A_{LV}$ and $A_{LS}$ denote the liquid-vapor and liquid-solid interface areas. For a droplet placed on a conical surface with half-apex angle, $\alpha$, we can use a finite element code, Surface Evolver[27], to deduce its shape from minimizing $U$ for a fixed droplet volume. The dots plotted in Fig. 3a show the resulting normalized surface free energy, $u = (\gamma A_s)^{-1} U$, versus $r_s \kappa$, where $r_s$ and $A_s = 4\pi r_s^2$ denote the radius and surface area of a spherical droplet of the same volume, $\kappa$ denotes the local curvature which is equal to $-R^{-1}\cos\alpha$ or $R^{-1}\cos\alpha$ when the droplet is placed on the external or internal conical surface with the local cross-section radius $R$. The results in Fig. 3a for various contact angles show the same trend for the surface free energy independent of whether the surface is hydrophilic or hydrophobic.

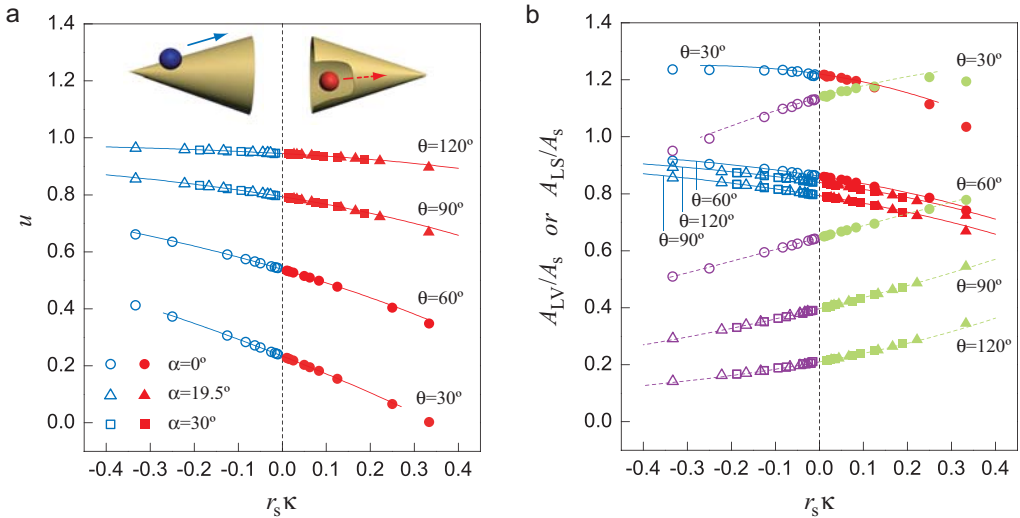

**Figure 3.** Droplet's free energy and interface areas as functions of curvature. **a,** Numerical results based on finite-element analysis for normalized total free energy $u$ of a small water droplet on a conical surface, as a function of local curvature, $\kappa$, directly below the center of mass of the droplet. Results for water contact angles $\theta = 120°, 90°, 60°$, and $30°$ are shown. For the droplet on the outer surface of the cone, this is a repulsive potential surface, relative to the cone apex, and for the droplet on the inside, it is attractive. The square and triangle symbols represent half-apex angles of the cone $\alpha = 30°$ and $19.5°$, respectively, and the circular dots represents cylinders ($\alpha = 0°$). **b,** The solid and dashed lines show how the liquid-vapor interface area $A_{LV}$ and the liquid-solid interface area $A_{LS}$ respectively vary with the local curvature.

The plotted results in Fig. 3a are scale-independent and are thus valid for different sized droplets and cones, from millimeters to nanometer. Furthermore, we find that the results are independent of half-apex angles, as numerically validated in Fig. 3a with $\alpha = 0°$ (circles for cylindrical tubes), $19.5^0$ (triangles), and $30^0$ (squares), and theoretically proved in SI.

The origin of curvi-propulsion can be understood by considering how $A_{LV}$ and $A_{LS}$ vary with the curvature $\kappa$, as illustrated by the solid and dashed lines respectively in Fig. 3b. The dominant energetic effect is that $A_{LV}$ decreases monotonically with $\kappa$. Since $A_{LS}$ increases with $\kappa$, from Eqn. 1 we see that this will either enhance or weaken the effect of decreasing $A_{LV}$ when the surface is hydrophilic ($\cos\theta > 0$) or hydrophobic ($\cos\theta < 0$), respectively. In both cases the net result, as shown in Fig. 3a, is that $U$ decreases monotonically with $\kappa$. Hence a droplet on the external cone surface moves towards $\kappa = 0$ (away from the cone apex) and on the internal cone surface towards large positive $\kappa$ (towards the cone apex). This holds true for both hydrophilic and hydrophobic surfaces, though the net force, and hence the maximum velocity, will be smaller in the hydrophobic case.

Although an analytical relationship between $u$ and $r_s\kappa$ may not exist, its tangent line equation at $r_s\kappa = 0$ can be shown to have the following form:

$$u = u_0 - \eta r_s \kappa, \qquad (2)$$

with

$$u_0(\theta) = \sqrt[3]{(2+\cos\theta)\sin^4\frac{\theta}{2}}, \quad \eta(\theta) = \frac{1+3\cos^2\theta}{4(2+\cos\theta)}\tan^{-2}\frac{\theta}{2}, \qquad (3)$$

where $u_0$ represents the normalized surface energy as a droplet is placed on a flat surface ($\kappa = 0$). Thus, a droplet on the cone surface is subject to a curvi-propulsion force, $F_c = -dU/ds$, that can be approximated when the droplet is relatively small ($r_s/R < 1$) as

$$F_c \approx \frac{3\gamma V}{2R^2}\eta(\theta)\sin 2\alpha, \qquad (4)$$

where $V$ is the droplet volume and $s$ is a meridian coordinate pointing to the curvature increasing direction.

Interestingly, the curvi-propulsion force $F_c$ scales as $R^{-2}$, analogous to an electrostatic force between two particles, with the effective potential between the droplet and the cone apex being repulsive on the outer surface, and attractive on the inner surface.

If dissipation were negligible, then a droplet placed at any position, characterized by radius $R_0$, on a conical surface would always start spontaneous motion. From Eqn. 4, we can obtain the explicit solution for the speed reached at position $R$ as follows:

$$v_{id}(R) = k v_c \left| \frac{\lambda_c}{R_0} - \frac{\lambda_c}{R} \right|^{1/2}, \qquad (5)$$

where $\lambda_c = (\gamma/\rho g)^{1/2}$ and $v_c = (\gamma g/\rho)^{1/4}$ are the capillary length and speed, respectively, that have the values $\lambda_c \approx 2.7$mm and $v_c \approx 163$mm/s for water at room temperature, and $k = [6\eta(\theta)\cos\alpha]^{1/2}$ is a monotonically decreasing function of $\theta$ with values of order of magnitude 1 for a large range of $\theta$, see Fig. 4a. We call $v_{id}$ in Eqn. 5 the *ideal* spontaneous moving speed, which is valid for droplets on both internal and external conical surfaces. On the external surface, the ideal speed has an upper limit $v_{up} = k v_c (\lambda_c/R_0)^{1/2}$ that is approachable as $R \to \infty$. Compared with the experimental and the numerical results shown in Figs. 1 and 2, the upper limits $v_{up}$ are equal to 1.1 m/s on the millimeter scale ($\theta \approx 41°$, $\Delta\theta \approx 4°$, $\alpha \approx 1°$, $r_s = 0.62$mm) and 610 m/s on the nanoscale ($\theta \approx 51°$, $\Delta\theta \approx 0°$, $\alpha \approx 19.5°$, $r_s = 1$ nm). These velocities are several times those observed, due to ignoring hysteresis and viscosity effects.

The hysteresis resistant force, $F_h$, can be related to the contact angle hysteresis approximately through the equation $F_h \approx \gamma w(\cos\theta_r - \cos\theta_a)$ [9], where $w$ is the width of the liquid-solid contact area, and $\theta_a$ and $\theta_r$ denote the advancing and receding contact angles. As $r_s/R \ll 1$, we can further derive

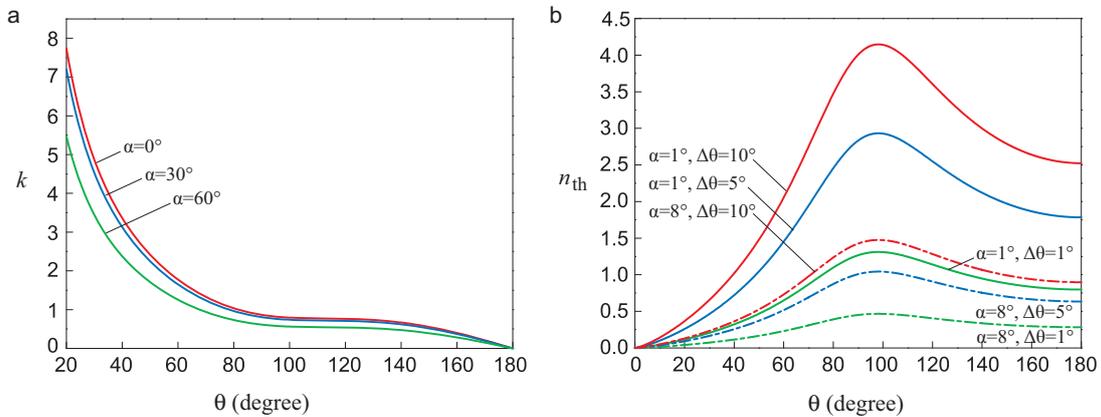

**Figure 4.** The main factors which characterize the spontaneous motion. **a**, The factor $k$ has values of order of magnitude 1 for a large range of $\theta$, and is almost invariant for small half-apex angle $\alpha$. **b**, The threshold $n_{th}$ which defines the minimal relative droplet's size $r_s/R$ that the droplet could be driven in motion by the curvi-propulsion. Much smaller droplets can spontaneously move on cone surfaces as the contact angles $\theta$ are small, compared with those as $\theta$ is near 98°.

$$F_h = \frac{3\gamma V}{2r_s^2} \frac{(\cos\theta_r - \cos\theta_a)\sin\theta}{\pi u_0(\theta)}.  \quad (6)$$

Comparing (4) with (6) yields

$$\frac{F_c}{F_h} = \left(\frac{r_s}{n_{th}R}\right)^2, \quad (7)$$

where we have denoted a threshold, $n_{th}$, for spontaneous motion as

$$n_{th} = \sqrt{\frac{\cos\theta_r - \cos\theta_a}{\pi\sin 2\alpha} \frac{\sin\theta}{\eta(\theta)u_0(\theta)}}. \quad (8)$$

In Fig. 4b we plot $n_{th}$ as a function of $\theta$ for some typical values of the hysteresis $\Delta\theta$ and half-apex angle $\alpha$ with $\theta_a$, $\theta_r = \theta \pm \Delta\theta/2$. From Eqn. 7, values of $r_s/R$ smaller or larger than the threshold $n_{th}$ correspond to $F_c > F_h$ or $F_c < F_h$, respectively. Consequently, any droplet placed on a conical surface at a position satisfying $R_0 < r_s/n_{th}$ will always start to move spontaneously. The maximum speed is attained for $R$ at the threshold. The value of the maximum speed is a factor $(1-n_{th}R_0/r_s)$ times the ideal speed limit $v_{up} = kv_c(\lambda_c/R_0)^{1/2}$. For the experiments on the glass cone, we can estimate the speed reduction due to $F_h$ to be about 20%.

Some remarkable implications that can be easily deduced from Fig. 4B are summarized below. (i) Slightly hydrophobic surfaces ($\theta = 98.2^0$) are the most difficult on which to achieve spontaneous motion, and surfaces with smaller or larger contact angles allow motion of smaller droplets. (ii) Making surfaces more hydrophilic is an effective way to either move smaller droplets, or achieve higher speeds for droplets with same volume, in agreement with the observations shown in Figures 1 and 2. (iii) Unlike the wettability gradient force that cannot move droplets smaller than certain critical size defined by the hysteresis, the curvi-propulsion force diverges for any small droplets sufficiently near the tip of a cone.

The curvi-propulsion mechanism presents two additional advantages over the wettability gradient mechanism. First, motion over much longer distances, $L = (n_{cr}\sin\alpha)^{-1}r_s$, can be achieved by curvi-propulsion. For example, using a cone with $\theta = 10°$, $\Delta\theta = 5°$, and $\alpha = 1°$, the transportable distance $L$ is more than one hundred times the droplet size $r_s$, while the wettability gradient mechanism can at most make a droplet run a distance about ten times of the droplet size (see Supplementary Information for details). Second, a cone surface is comparably easier to fabricate and chemically more stable than a wettability gradient surface, particularly at small scale.

Finally, we provide a simple analysis on dissipation effect due to viscosity. Using a similar analysis to that given in Ref. 18, we can obtain an approximation $F_v = Cv$ with $C \approx \mu l(1+\cos\theta)r_s/u_0$ for moving droplets on a cone surface, where $\mu$ is the viscosity and $l$ is a logarithmic factor which is a constant of order 15 or 5 for a surface that is either dry or pre-wetted[18,28,29]. This analysis gives the following estimate:

$$F_v = \frac{3\gamma V}{2r_s^2} \frac{(1+\cos\theta)l}{2\pi u_0(\theta)} \frac{v}{v_\mu}, \qquad (9)$$

where the capillary-viscosity velocity $v_\mu = \gamma/\mu$ has the value about 73 m/s for water at room temperature. For the millimeter-scale experiments on the glass cone, we have $F_v/F_h \approx 300v/v_\mu$. Thus, $F_v$ and $F_h$ at the millimeter scale would be the same order. At the nanoscale, however, the curvi-propulsion can instantly drive water droplets into speeds that are comparable with $v_\mu$ (see Fig. 2), the viscosity resistance thus becomes dominant. The dot-dashed line in Fig. 2b shows some predicted results that agree qualitatively well with observation.

In conclusion, we have demonstrated ultra-high velocity motion of droplets moving on conical surfaces, by experiment on the sub-millimeter scale and by molecular dynamics simulation on the nanoscale. The peak velocities achieved this way can exceed 100m/s on the nanoscale, orders of magnitude faster than any previously reported result for spontaneous droplet motion. We explain this remarkable effect, which we term curvi-propulsion, as being due to the much higher curvature gradients that nanoscale droplets can experience. We corroborate the effect using finite-element analysis and simple scaling arguments.

If suitable substrates can be designed to exploit this phenomenon, for example using arrays of nanoscale conical structures, we speculate that curvi-propulsion could be useful for a range of practical applications where mass transport via droplet motion plays a key role, including rapid cooling, passive water collection and micro chemical synthesis. The curvi-propulsion mechanism may be also useful in understanding inter-cell transport through cytonemes or filopedial bridges[30].

**Acknowledgements:** Financial support from the NSFC under grant No.10872114, No.10672089, and No. 10832005 is gratefully acknowledged. The authors greatly thank Professor David Quéré for a critical reading of the manuscript.

# Supplementary Information

For "Driving Droplets by Curvi-Propulsion"

by Cunjing Lv, Chao Chen, Yin-Chuan Chuang, Fan-Gang Tseng, Yajun Yin, Francois Grey, Quanshui Zheng

**S.1 Details of Experiments**

**Materials:**
Glass capillaries with tip/end diameters of 140um/1.5mm were used. The tips of the capillaries were sharpened by a flaming/brown micropipette puller (Model P-1000, Sutter Instrument, U.S.A.) to obtain needle-shape capillaries.

Anhydrous alcohol (99.5%) and anhydrous toluene (99.9%) were purchased from Shimakyu's Pure Chemical (Osaka, Japan) and J. T. Baker (Phillipsburg, NJ, U.S.A). MTS (Methyltrichlorosilane, $CH_3SiCl_3$, 99% purity, catalogue number: M85301) was purchased from Sigma-Aldrich, USA.

**Methods:**
1. *Surface cleaning*

   Acetone and IPA were used to preclean surfaces of the glass capillaries. Then deionized (DI) water with a resistance of 20MΩ was used to carefully clean the surfaces.

2. *Surface Modification process*

   Outer surfaces of all the capillaries were pretreated by oxygen plasma at 100 W and 75 mtorr chamber pressure for 300 s. Then the capillaries were put in 0.014 M MTS solution in anhydrous toluene for 75 min under 23°C and 75 % RH environmental conditions. For removing the residual molecules that were not immobilized from the surface, the capillaries were then rinsed in anhydrous toluene, ethanol, a mixed solution of ethanol and DI (1:1), and DI water in a sequence. Compressed dry air was employed to blow dry the capillaries, and finally the capillaries were put into an oven for annealing at 120°C for 10 min [1].

3. *Measurement*

   The static contact angle (CA) and contact angle hysteresis (CAH) of each surface before and after treatment were measured with FTA 200 instrument (First Ten Angstroms, U.S.A.) by using pipette to drop 2 $\mu$L DI water on a flat glass slide (Kimble glass, Owen-Illinois, U.S.A.) in stead of the glass capillaries because of the difficulties in defining contact angle measurement on a curved surface.

   Images of the moving 1 $\mu$L droplet were captured with a high speed camera (Fastcam-Ultima APX, Photron) coupled with a long focus optical system (Zoom 700X with internal focus vertical illuminator, OPTEM International U.S.A).

**Results & Discussions:**

After basic clean, the plain glass capillary surface showed a CA of 41° and CAH of 4°. Then it became highly hydrophilic with a CA of 5° and CAH of 2° after oxygen plasma treatment.

The capillary surface was modified with a 3D nano-textures of MTS aggregations[1,2] and treated by oxygen plasma. It finally presented a CA of 0° and CAH of 1.5°. (as shown in Fig. S1)

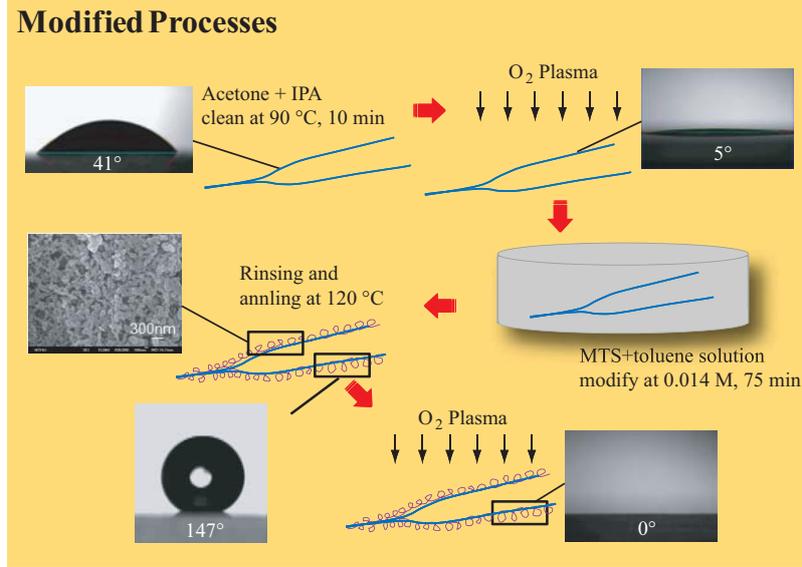

**Fig. S1.** Chemical modification processes of the capillary surface.

## S.2 Theory

### S.2.1 Relationship between surface energy of water droplet with curvature

The liquid-vapor interface energy, $\gamma$, specifies two parameters, the capillary length $\lambda_c = (\gamma/\rho g)^{1/2}$ and capillary speed $v_c = (\gamma g/\rho)^{1/2}$, where $\rho$ is the liquid mass density and $g$ is the gravity constant. For water at the room temperature, we have $\lambda_c \approx 2.7$mm and $v_c \approx 163$mm/s.

For a droplet placed on a solid surface, if the droplet size is smaller than the capillary length, we can ignore the gravity influence and express the total free energy as:

$$U = A_{LV}\gamma + A_{LS}(\gamma_{LS} - \gamma_{SV}) = \gamma(A_{LV} - A_{LS}\cos\theta) \quad (S1)$$

where $A_{LV}$ and $A_{LS}$ denote the liquid-vapor and liquid-solid interface areas, $\gamma_{LS}$ and $\gamma_{SV}$ the liquid-solid and solid-vapor interface energies, and $\theta$ is the Young's contact angle which is associated with the interface energies in the form $\cos\theta = (\gamma_{SV} - \gamma_{LS})/\gamma$.

For a given volume, a droplet may have various shapes. However, the real one in static is such that the total free energy $U$ is the minimum. For a droplet placed on a flat sphere surface, the droplet's shape is a spherical cap. For a droplet placed on a conical surface, it is hard to find the analytical solution of the shape. Hereinafter we show that we can find an analytical approximate.

From the theory of differential geometry[3], it is known that locally the departure of a curved surface from it tangent plane is determined by the two principal curvatures of the surface, $\kappa_1$ and $\kappa_2$. Therefore, if the droplet's size is smaller than the curvature radii $1/|\kappa_1|$, $1/|\kappa_2|$, then the free energy $U$ depends only upon the local property of the surface, namely $\kappa_1$ and $\kappa_2$: $U = U(\kappa_1,\kappa_2)$. Mathematically, if denoting by $r_s$ the radius of a spherical droplet with the same volume, we can write the relatively small droplet condition as $r_s\kappa \ll 1$, where $\kappa = \max\{|\kappa_1|,|\kappa_2|\}$. Thus, we can take the Taylor expansion of $U(r_s\kappa_1,r_s\kappa_2)$ as follows:

$$U(r_s\kappa_1,r_s\kappa_2) = U(0,0) + \frac{\partial U(0,0)}{\partial(r_s\kappa_1)}r_s\kappa_1 + \frac{\partial U(0,0)}{\partial(r_s\kappa_2)}r_s\kappa_2 + O((r_s\kappa)^2). \qquad (S2)$$

Since water is isotropic, it yields

$$\frac{\partial U(0,0)}{\partial(r_s\kappa_1)} = \frac{\partial U(0,0)}{\partial(r_s\kappa_2)}. \qquad (S3)$$

Consequently,

$$U(r_s\kappa_1,r_s\kappa_2) = U(0,0) + \frac{\partial U(0,0)}{\partial(r_s\kappa_1)}2r_s H + O((r_s\kappa)^2), \qquad (S4)$$

where $H = (\kappa_1 + \kappa_2)/2$ is the mean curvature. In other words, if we ignore a higher order small quantity, then $U$ depends upon only the mean curvature, rather than the Gauss curvature $K = \kappa_1\kappa_2$.

Based on the above result, we can first find the analytical solution of $U(R)$ for droplets placed on a spherical surface with radius $R$, we then replace $R^{-1}$ by the mean curvature $H$ for other curved surfaces. In particular, for a cone with the half apex angle $\alpha$ and local cross-section radius $R$, the two principal curvatures are $\kappa_1 = R^{-1}\cos\alpha$ and $\kappa_2 = 0$, respectively. We can replace $R^{-1}$ in $U(R)$ for the sphere with $H = (R^{-1}\cos\alpha)/2$ for the cone.

**S.2.2 Surface energy of water droplet on a spherical surface**

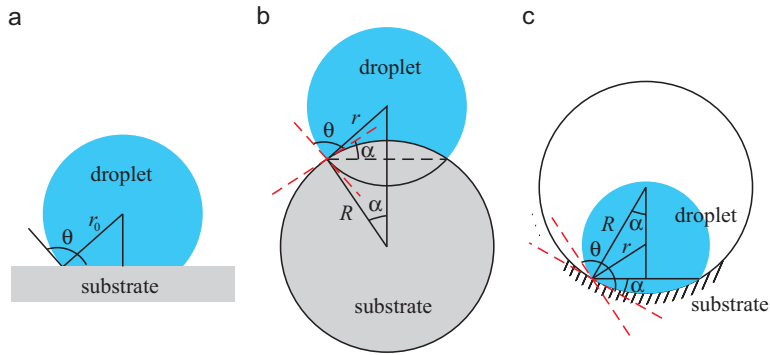

**FIG. S2.** Sketch about the wetting states of the same water droplet on a flat **a**, outer **b** and inner **c** of a curved surfaces, respectively.

As the droplet is placed on a flat surface (Fig. S2a), the free energy can be easily solved as:

$$U_0 = \pi r_0^2 (2 - 3\cos\theta + \cos^3\theta)\gamma = \frac{3V\gamma}{r_0}, \tag{S5}$$

where

$$\begin{aligned} V &= \frac{\pi}{3} r_0^3 \left(2 - 3\cos\theta + \cos^3\theta\right) \\ &= \frac{\pi}{3} r_0^3 (1-\cos\theta)^2 (2+\cos\theta), \\ &= \frac{4\pi}{3} r_0^3 (2+\cos\theta)\sin^2\frac{\theta}{2} \end{aligned} \tag{S6}$$

is the droplet volume, and $r_0$ is the radii of the water-solid contact area. Denoting by $r_s$ the radius of a spherical droplet of the same volume $V$, from (S6) we can further have

$$\frac{r_s}{r_0} = \sqrt[3]{\frac{2-3\cos\theta+\cos^3\theta}{4}} = \sqrt[3]{(2+\cos\theta)\sin^4\frac{\theta}{2}}. \tag{S7}$$

When water droplet with the same volume is placed on the external spherical surface (Fig. S2b), referred to (S6) and Fig. S2b we can get the following relationships:

$$\begin{cases} V = \dfrac{\pi}{3} r^3 (1-\cos\hat{\theta})^2 (2+\cos\hat{\theta}) - \dfrac{\pi}{3} R^3 (1-\cos\alpha)^2 (2+\cos\alpha), \\ R\sin\alpha = r\sin\hat{\theta}, \end{cases} \tag{S8}$$

where $r$ is the radius of the water droplet on the spherical surface, $R$ is the radius of the contact area, $\hat{\theta} = \theta + \alpha$, and $\alpha$ as shown in (Fig. S2b) is the half-spanned angle of the wetted area of the sphere. The two interface areas are:

$$A_{LV} = 2\pi r^2 (1-\cos\hat{\theta}), \quad A_{LS} = 2\pi R^2 (1-\cos\alpha), \tag{S9}$$

From Eqns. (S1), (S8), and (S9) we can numerically solve $U$ as a function of the means curvature, plotted as the smooth solid and dotted lines in Fig. 3ab. The excellent agreement with the numerical results for cone surfaces proves that ignoring the dependence of $U$ upon the Gauss curvature would be an excellent approximation. Similar results can be obtained as the water droplet is placed on the internal surface of the sphere (see Fig. S2c).

**S.2.3 Linear relationship between mean curvature $\kappa$ and surface energy $U$**

As $\alpha \ll 1$, by ignoring higher order small quantities of $\alpha$ it is not difficult to get:

$$\frac{r}{r_0} \approx 1 - \frac{3(1+\cos\theta)\sin\theta}{4(1-\cos\theta)(2+\cos\theta)}\alpha, \tag{S10}$$

$$V \approx \frac{\pi}{3} r^3 \left[ \left(2 - 3\cos\theta + \cos^3\theta\right) + 3\alpha \sin^3\theta \right] - \frac{\pi}{4}\alpha^4 R^3, \tag{S11}$$

$$A_{LV} \approx 2\pi r^2 \left(1 - \cos\theta + \alpha\sin\theta\right),$$

$$A_{LS} = 2\pi R^2 \left(1 - \cos\alpha\right) \approx \pi\alpha^2 R^2. \tag{S12}$$

Introduce the normalized free energies

$$u(\theta) = \frac{U}{\gamma A_s}, \quad u_0(\theta) = \frac{U_0}{\gamma A_s}, \tag{S13}$$

where $A_s = 4\pi r_s^2$ denote the surface energy of a spherical droplet of the same volume. From Eqns. (S5) and (S7) we have

$$u_0(\theta) = \sqrt[3]{(2+\cos\theta)\sin^4\frac{\theta}{2}}, \tag{S14}$$

From Eqns. (S1), (S10-S12), we can get the tangent relation of the normalized surface free energy as follows:

$$u = u_0 + 2\eta\left(\frac{r_s}{R}\right), \tag{S15}$$

where

$$\eta(\theta) = \frac{1 + 3\cos^2\theta}{4(2+\cos\theta)} \tan^{-2}\frac{\theta}{2}. \tag{S16}$$

Since for droplet placed on the external spherical surface the mean curvature is equal to $-1/R$, for cone we should replace $-1/R$ in Eqn. (S15) by the mean curvature $\kappa/2$. The above analysis finally results in:

$$u = u_0 - \eta r_s \kappa. \tag{S17}$$

**S.2.4 Transportable length**

As shown by Eqn. (7), a droplet would start a spontaneous motion on a conical surface whenever $r_s/R < n_{th}$. Since $R = L \tan\alpha$, the length $L$ from the cone tip valid for starting spontaneous motion is equal to

$$L = \frac{r_s}{n_{th} \tan\alpha}. \tag{S18}$$

For $\theta = 10°$, $\Delta\theta = 5°$, and $\alpha = 1°$, we have $n_{th} = 0.1$ and $L = 570 r_s$.

As illustrated in Fig. S3a, a droplet may start a spontaneous motion in the $x$-direction on a flat surface with contact angle gradient $\theta'(x) < 0$ if and only if the advancing contact angle $\theta_a$ at the leading edge $x+r_0$ is smaller than the receding contact angle $\theta_r$ at the trailing edge $x - r_0$, where $r_0$ is the radius of the liquid-solid contact area. Using Taylor expansion, we have

$$\theta_a(x+r_0) - \theta_r(x-r_0) \approx \theta_a(x) - \theta_r(x) + [\theta_a'(x) + \theta_r'(x)]r_0$$
$$\approx \Delta\theta(x) + \bar{\theta}'(x)2r_0 \tag{S19}$$

where $\bar{\theta} = (\theta_a + \theta_r)/2$ is the mean contact angle, $\Delta\theta = \theta_a - \theta_r$ is the contact angle hysteresis, and ()' denotes derivative d()/d$x$. Therefore, the condition of allowing spontaneous motion is

$$2r_0 > \frac{\Delta\theta}{-\bar{\theta}'}. \tag{S20}$$

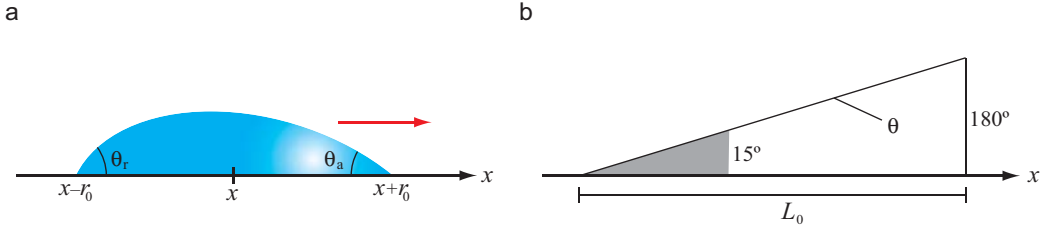

**Fig. S3.** a, Droplet moving toward a lower contact angle region. b, Contact angle gradient region and transportable region (grey).

Using Eqns. (7) and (S14), we further have

$$\frac{2r_s}{u_0(\theta)} > \frac{\Delta\theta}{-\bar{\theta}'}. \tag{S21}$$

For an estimate, we assume $\bar{\theta} = \theta$, $\Delta\theta = 5°$, and $\bar{\theta}' = -\pi/L_0$. For $2r_s/L_0 = 1/100$, to satisfy the condition shown in Eqn. (S21), the contact angle must be lower than $\theta < 15°$, as illustrated in Fig. S3b. Therefore, the transportable length is less than $(15°/180°) L_0 = 17r_s$.

### S.3 Supplementary Movies

**Movie 1 (S1.mov, 1.74MB) –Side view of 1μL water droplet move spontaneously on the coincal surface treated with MTS nanotextured plus $O_2$ plasma.**

When a 1.0 μL water droplet is released gently enough from the syringe, it will move spontaneously on the conical surface (see Fig. 1a). Glass capillaries with tip/end diameters of 140um/1.5mm and surface condition is MTS nanotextured plus $O_2$ plasma treated with $\theta \approx 0°$, $\Delta\theta \approx 1.5°$. The actual speed of the water droplets in experiment is 200 times higher than which is shown in the videos.

**Movie 2 (S2.mov, 273KB) –Side view of 1μL water droplet move spontaneously on the conical surface treated with $O_2$ plasma.**

When a 1.0 $\mu$L water droplet is released gently enough from the syringe, it will move spontaneously on the conical surface. The surface condition is $O_2$ plasma treated with $\theta \approx 5°$, $\Delta\theta \approx 2°$. The actual speed of the water droplets in experiment is 10 times of which shown in the video.

**Movie 3 (S1.mov, 498KB) – Side view of 1$\mu$L water droplet move spontaneously on the plain coincal surface.**

When a 1.0 $\mu$L water droplet is released gently enough from the syringe, it will move spontaneously on the conical surface. The surface condition is plain with $\theta \approx 41°$, $\Delta\theta \approx 4°$. The actual speed of the water droplets in experiment is 10 times of which shown in the videos.

**Movie 4 (S4.mov, 3.36MB) – Molecular dynamics simulation results of water droplet self-motion**

The water droplet on the outer/inner conical surface is spontaneously moving toward the larger/smaller cross-section area regardless of the value of the water contact angle ($\theta$=50.7° (hydrophilic), 138° (hydrophobic), or nearly 180° (superhydrophobic)). The droplet is bouncing back when it meets the boundary.